\documentclass{ws-mpla}
\pdfoutput=1
\usepackage[super]{cite}
\usepackage{graphicx}
\usepackage{amssymb}
\usepackage{amsmath}
\usepackage{amsmath,multicol}
\usepackage{tabularx,tabulary}
\usepackage{bm}
\usepackage{comment}
\usepackage{caption}
\usepackage{subcaption}
\usepackage{xcolor}
\def\a {\epsilon}
\def\g {\gamma}
\def\A  {{\cal A}}
\def\I  {{\cal I}}
\def\II {{\cal J}}

\def\P  {{\cal P}\!}
\def\Q  {{\cal Q}_{cc}}
\def\R  {{\cal R}}

\def\J  {$J/\psi$ }
\def\ds {$D_s$ }

\def\ap#1#2#3   {{\rm Ann. Phys. (NY)}       #1 (#3) #2}
\def\apj#1#2#3  {{\rm Astrophys. J.}         #1 (#3) #2}
\def\apjl#1#2#3 {{\rm Astrophys. J. Lett.}   #1 (#3) #2}
\def\app#1#2#3  {{\rm Acta. Phys. Pol.}      #1 (#3) #2}
\def\chp#1#2#3  {{Chin.\ Phys. }             #1 (#3) #2}
\def\cpc#1#2#3  {{\rm Computer Phys. Comm.}  #1 (#3) #2}
\def\dum#1#2#3  {{~}                         #1 (#3) #2}
\def\epjc#1#2#3 {{\rm Eur. Phys. J. C}       #1 (#3) #2}
\def\err#1#2#3  {{\it Erratum}               #1 (#3) #2}
\def\ib#1#2#3   {{\it ibid.}                 #1 (#3) #2}
\def\jcp#1#2#3  {{\rm J. Comp. Phys.}        #1 (#3) #2}
\def\jmp#1#2#3  {{\rm J. Math. Phys.}        #1 (#3) #2}
\def\jhep#1#2#3 {{\rm JHEP}                  #1 (#3) #2}
\def\ijmp#1#2#3 {{\rm Int. J. Mod. Phys.}    #1 (#3) #2}
\def\jpg#1#2#3  {{\rm J. Phys. G.}           #1 (#3) #2}
\def\mpl#1#2#3  {{\rm Mod. Phys. Lett.}      #1 (#3) #2}
\def\mpla#1#2#3 {{\rm Mod. Phys. Lett. A}    #1 (#3) #2}
\def\nat#1#2#3  {{\rm Nature (London)}       #1 (#3) #2}
\def\ncim#1#2#3 {{\rm Nuovo Cimento}         #1 (#3) #2}
\def\nca#1#2#3  {{\rm Nuovo Cimento A}       #1 (#3) #2}
\def\ncb#1#2#3  {{\rm Nuovo Cimento B}       #1 (#3) #2}
\def\nim#1#2#3  {{\rm Nucl. Instr. Meth.}    #1 (#3) #2}
\def\njp#1#2#3  {{New J. Phys. }             #1 (#3) #2}
\def\np#1#2#3   {{\rm Nucl. Phys.}           #1 (#3) #2}
\def\npb#1#2#3  {{\rm Nucl. Phys. B}         #1 (#3) #2}
\def\pan#1#2#3  {{\rm Phys. At. Nuclei}      #1 (#3) #2}
\def\pl#1#2#3   {{\rm Phys. Lett.}           #1 (#3) #2}
\def\plb#1#2#3  {{\rm Phys. Lett. B}         #1 (#3) #2}
\def\prep#1#2#3 {{\rm Phys. Rep.}            #1 (#3) #2}
\def\prev#1#2#3 {{\rm Phys. Rev.}            #1 (#3) #2}
\def\prc#1#2#3  {{\rm Phys. Rev. C}          #1 (#3) #2}
\def\prd#1#2#3  {{\rm Phys. Rev. D}          #1 (#3) #2}
\def\prev#1#2#3 {{\rm Phys. Rev.}            #1 (#3) #2}
\def\prl#1#2#3  {{\rm Phys. Rev. Lett.}      #1 (#3) #2}
\def\prs#1#2#3  {{\rm Proc. Roy. Soc.}       #1 (#3) #2}
\def\ptp#1#2#3  {{\rm Prog. Theor. Phys.}    #1 (#3) #2}
\def\ptep#1#2#3 {{\rm Prog. Theor. Exp. Phys.}  #1 (#3) #2}
\def\ps#1#2#3   {{\rm Physica Scripta}       #1 (#3) #2}
\def\rmp#1#2#3  {{\rm Rev. Mod. Phys.}       #1 (#3) #2}
\def\rpp#1#2#3  {{\rm Rep. Prog. Phys.}      #1 (#3) #2}
\def\sa#1#2#3   {{Sci.\ Acta }               #1 (#3) #2}
\def\sci#1#2#3  {{Science }                  #1 (#3) #2}
\def\sjnp#1#2#3 {{\rm Sov. J. Nucl. Phys.}   #1 (#3) #2}
\def\spj#1#2#3  {{\rm Sov. Phys. JETP}       #1 (#3) #2}
\def\spu#1#2#3  {{\rm Sov. Phys.-Usp.}       #1 (#3) #2}
\def\yaf#1#2#3  {{\rm Yad. Fiz.}             #1 (#3) #2}
\def\zp#1#2#3   {{\rm Zeit. Phys.}           #1 (#3) #2}
\def\zpa#1#2#3  {{\rm Zeit. Phys. A}         #1 (#3) #2}
\def\zpc#1#2#3  {{\rm Zeit. Phys. C}         #1 (#3) #2}
\def\et         {{\it et al.}}
\newcommand{\n}{\nonumber \\}
\begin{document}

\markboth{A. G. Bagdatova, S. P. Baranov, A. S. Sakharov}{Study of exclusive
two-body $W$ decays with fully reconstructible kinematics}

\catchline{}{}{}{}{}

\title{Study of exclusive two-body $W$ decays with fully reconstructible kinematics}
\author{Alsu G. Bagdatova$^{\ast}$, ~{\rm Sergey P. Baranov}$^{\dagger}$}
\address{P.N. Lebedev Institute of Physics,
              53 Lenin Avenue, 119991 Moscow, Russia\\
$^{\ast}$~bagdatovaag@lebedev.ru\\
$^{\dagger}$~baranovsp@lebedev.ru}
\author{Alexander\ S.\ Sakharov}
\address{Physics Department, Manhattan College, Manhattan College Parkway,
Riverdale, NY 10471, United States of America and\\
Experimental Physics Department, CERN, CH-1211 Gen\`eve 23, Switzerland\\
Alexandre.Sakharov@cern.ch}

\maketitle

\pub{Received (Day Month Year)}{Revised (Day Month Year)}

\begin{abstract}
In the framework of electroweak theory and perturbative quantum chromodynamics,
we examine various exclusive decay channels of $W$ bosons that can be fully or
partially reconstructed. Our findings provide predictions for the partial widths
and address some gaps in previous literature. We also place a strong emphasis
on understanding and estimating the associated theoretical uncertainties.

\end{abstract}

\ccode{PACS Nos.: 12.38.Bx, 14.70.Fm, 13.38.Be, 12.39.Jh}


\section{Introduction}

Rare hadronic decays of W bosons
are discussed as
having the potential to offer a new method for measuring the W boson mass
through visible decay products at future colliders.
An example of such a decay is $W\to J/\psi\,D_s$ with $J/\psi\to l^+l^-$
and $D_s\to~K^+K^-\pi$; here $J/\psi\to l^+l^-$ provides the necessary
trigger signature. This decay, along with a wider class of decays
$W\to\Q +D_s^{(*)}$ where $\Q$ can be any quarkonium state such as
$J/\psi$, $\eta_c$, $\psi'$, $\chi_{c0}$, $\chi_{c1}$, $\chi_{c2}$, $h_c$,
has been theoretically considered in Ref.~\cite{LuchLik}.

Radiative decays, such as $W\to D_s\g$ or $W\to D_s^*\g$, have the
potential to test the Standard Model and, probably, uncover new physics
beyond the Standard Model, as they involve the three-boson coupling
vertex $WW\g$~\cite{Baur}. Theoretical calculations for these decays
can be found in references~\cite{Parsa,Grossman}. The aim of these
studies was to determine the feasibility and accuracy of observing these
decay modes at current and future particle accelerators~\cite{Grossman}.

The above cited works are important and provide valuable insights,
however, they have certain limitations. Our aim is to address these
limitations in this note. The analysis presented in Refs.~\cite{Parsa,Grossman}
does not include decays into vector mesons  $W\to D_s^*\g$, and is
restricted only to the Light Cone (LC) technique. A comparison with
Nonrelativistic Quantum Chromodynamics (NRQCD) would provide a more
comprehensive picture. Ref.~\cite{LuchLik} provides an incomplete
analysis by ignoring the dominant contributions to the decays
$W\to J/\psi\,D_s^{(*)}$.

Our objective is to fill these gaps. Additionally, we aim to
examine the numerical stability of the calculations, which has
not been explored in previous publications. This involves
examining the sensitivity of the results to the choice of input
parameters, which can provide a deeper understanding of the
reliability of the calculations.

The rest of the paper is organized as follows.
In Sec. 2, we explain the technical details of our calculation.
In Sec. 3, we present and discuss the results.
Our findings are briefly summarised in Sec. 4.

\section{Calculation}
The list of processes considered in our note is:
\begin{eqnarray}
W &\to& D_s+\gamma  \label{c1}   \\
W &\to& D_s^*+\gamma \label{c2}  \\
W &\to& J/\psi+D_s     \label{d1} \\
W &\to& J/\psi+D_s^*   \label{d2} \\
W &\to& \psi(2s)+D_s   \label{d3} \\
W &\to& \psi(2s)+D_s^* \label{d4} \\
W &\to& \chi_c+D_s     \label{d5} \\
W &\to& \chi_c+D_s^*   \label{d6} \\
W &\to& B_c+B_s       \label{b1} \\
W &\to& B_c^*+B_s     \label{b2} \\
W &\to& B_c+B_s^*     \label{b3} \\
W &\to& B_c^*+B_s^*   \label{b4}
\end{eqnarray}

The above processes are supposed to be detected via the decay chains
~$J/\psi~\to~\mu^+~\mu^-$, ~$\psi'\to\mu^+\mu^-$, ~$D_s\to K^+K^-\pi$,
~$\psi'\to J/\psi\,\pi\,\pi$, ~$\chi_c\to J/\psi\,\gamma$,
~$D_s^*\to D_s\,\pi^0$, ~$D_s^*\to D_s\,\gamma$.

\begin{figure}[!ht]
\centerline{\includegraphics[width=15.0cm]{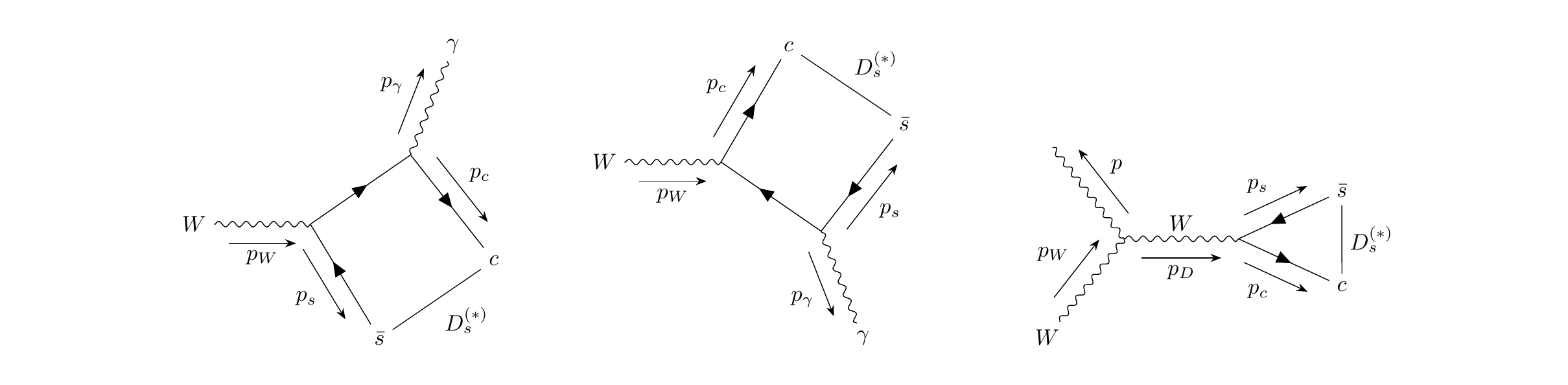}}
\centerline{\includegraphics[width=12.0cm]{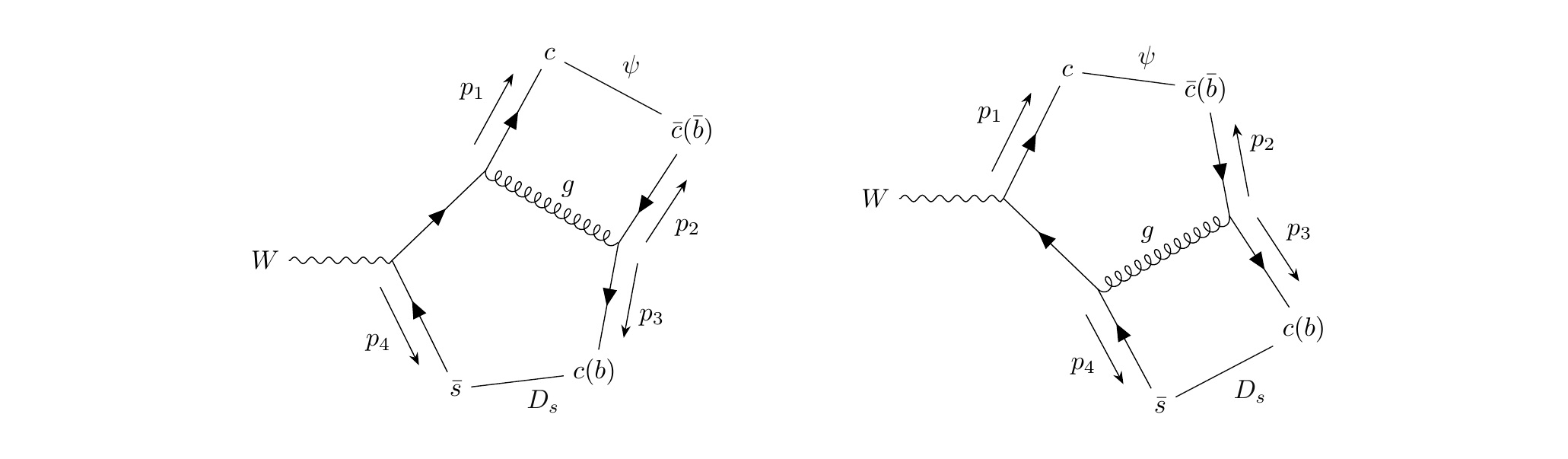}}
\vspace*{8pt}
\caption{Feynman diagrams describing radiative (upper row) and two-body
mesonic (lower row) decays of W boson.}
\protect\label{fig:FeynDiag}
\end{figure}

The calculation is based on the standard electroweak theory and perturbative QCD.
The corresponding Feynman diagrams are displayed in Fig.~\ref{fig:FeynDiag}. Large
energy release justifies the applicability of perturbative expansion. The relative
momentum of the decay products is large enough to make the final state interaction
negligible~\footnote{
This may be not fully true if we accept Light Cone (LC) model for the formation
of mesons. See our further discussion in Sec. 3 on the importance of the small
quark momentum region.}
thus validating the QCD factorization. Therefore, the formation of the final state
mesons can be described in terms of color-singlet wave functions. The details of
the relevant technique are explained in Refs.~\cite{Chang,Jones,Baier,Krase,Guber}.

The structure of the $W\to D_s^{(*)}\,\g$ decay amplitudes is
\begin{eqnarray}
&\A_1 =  C\,\a_W^\nu\,tr\Bigl\{ W_\nu\,{\P}_D\,\not{\!}\!\a_{\g}\,
\displaystyle{
\frac{(\not{\!}\!p_W-\not{\!}\!p_s+m_c)}{(p_W{-}p_s)^2{-}m_c^2}\Bigr\}},\nonumber \\
&\A_2 = C\,\a_W^\nu\,tr\Bigl\{ W_\nu\,
\displaystyle{
\frac{(\not{\!}\!p_c-\not{\!}\!p_W+m_s)}{(p_c{-}p_W)^2{-}m_s^2}\,
\not{\!}\!\a_{\g}\,{\P}_D\Bigr\}} ,\label{amp_gamD} \\
&\A_3 = C\,\a_W^\nu\,G_{\nu\mu\lambda}\,
 \a_{(\g)}^{\mu}\,\bigl[ g^{\lambda\sigma}-p_D^\lambda\,p_D^\sigma/m_W^2 \bigl]
 tr\bigl\{ W_\sigma\,{\P}_D\bigr\}/(m_D^2-m_W^2), \nonumber
\end{eqnarray}
where $C=\sqrt{3}\,e_q\,g_W$ includes the color factor and the coupling constants;
$g_W~=~eV_{cs}/\sqrt{8}\sin\theta_W$; and $e_q=2/3,\,-1/3,$ and 1 for $q=c,\,s,$ and
$W$. We follow the argumentation of Ref.~\cite{Grossman} pointing out that the
existence of triangle anomaly does not play essential role%
\footnote{It has been suggested in Ref.~\cite{Mangano} that the triangle anomaly could
produce a huge enhancement of the decay rates for $W\to D_s\g$, in analogy to
the case of $\pi^0\to\g\g$ amplitude. However, the present situation is rather
different. A careful inspection shows Ref.~\cite{Grossman} that the anomaly does not
exhibit a pole but is instead proportional to $1/m_W^2$.}.

The structure of the $W\to J/\psi+D_s^{(*)}$ decay amplitudes is
\begin{eqnarray}
{\A'}_4 & = & C\,\a_W^\nu\,\frac{1}{(p_{2} + p_{3})^2}\,tr\Bigl\{ W_\nu\,{\P}_D\,
\g^\mu\,{\P}_{\psi}\,\g_\mu\,\displaystyle{
\frac{(\not{\!}\!p_W-\not{\!}\!p_4+m_c)}{(p_W{-}p_4)^2{-}m_c^2}\Bigr\}} ,\nonumber\\
{\A'}_5 & = & C\,\a_W^\nu\,\frac{1}{(p_{2} + p_{3})^2}\,tr\Bigl\{ W_\nu\,
\displaystyle{
\frac{(\not{\!}\!p_1-\not{\!}\!p_W+m_s)}{(p_1{-}p_W)^2{-}m_s^2}}\,
\g^\mu\,{\P}_D\,\g_\mu\,{\P}_{\psi}\Bigr\},
\label{amp_psiD}
\end{eqnarray}
where $C=4/3\,g^2\,g_W$ and $g$ is the strong coupling charge.
The strong coupling constant is parametrized as
\begin{equation}
\alpha_s(\mu^2) = \frac{4\pi}{b_0~\ln(\mu^2/{\Lambda_{QCD}}^2)},
\end{equation}
$b_0 = 11 - {2/3}~n_f, \, n_f$ is the number of flavours ($n_f = 3$ for the decays
into charmed modes, and $n_f = 4$ for $b$-flavored modes).
The choice of the renormalization scale is dictated by the gluon
virtuality. So, we set $\mu^2 = (p_2+p_3)^2$.

In the expressions~(\ref{amp_psiD}),
$\a_W$, $\a_{\g}$ and $\a_{\psi}$ are the $W$, photon and \J polarization vectors;
$p_c$, $p_s$, $p_{\psi}$ and $p_D$ are the quark and meson momenta;
$m_c$, $m_s$, $m_{\psi}$ and $m_D$ the respective masses;
$W_\nu {=} \g_\nu\,(1{-}\g_5)$ is the standard $W$ boson to quark coupling; and
$G_{\nu\mu\lambda}$ is the Standard Model three-boson coupling vertex:
\begin{eqnarray}
G_{\nu\mu\lambda} &= (k_1-k_2)_\mu\,g_{\nu\lambda} + (k_2-k_3)_\nu\,
g_{\lambda\mu}+ (k_3-k_1)_\lambda\,g_{\mu\nu},
\end{eqnarray}
where $k_i$ denote the incoming boson 4-momenta.
Note that the amplitudes (\ref{amp_gamD}) do also contribute to the decays
(\ref{d1})-(\ref{d4}) through the photon conversion $\gamma^*\to J/\psi$.
The amplitude conversion factors read
\begin{eqnarray}
{\A'}(\g*\to\psi) &=& e_c\sqrt{3/\pi}\,\R_{\psi}(0)/m_\psi^{3/2}\\ \label{ampfactor}
{\A'}(\g*\to\psi) &=& e_c\,f_\psi/m_\psi
\end{eqnarray}
for the NRQCD and LC schemes, respectively (these schemes will be explained a bit later),
and the polarization vector $\a_{\g}$ has to be replaced with $\a_{\psi}$.
The amplitudes (\ref{amp_gamD}) extended with the $\gamma^*\to J/\psi$ conversion will
be referred to as ${\A'}_1,\,{\A'}_2,\,{\A'}_3$.

The diagrams (\ref{amp_gamD}) and (\ref{amp_psiD}) constitute two independent gauge
invariant sets. The interference between them is automatically taken
into account as we sum the amplitudes, not the squares (see eqs.(25), (26), (29)-(32)).

When calculating the amplitudes we use spin projector operators onto the pseudoscalar
(spin-singlet) and vector (spin-triplet) states, which guarantee that the $(c\bar{s})$
and $(c\bar{c})$ states have the intended quantum numbers:
\begin{eqnarray}
{\P}_D &=& \g_5\,(\not{\!}\!p_D+m_D)/2m_D^{1/2},\\
{\P}_{D^*}\!\!\! &=& \not{\!}\!\a_{D}\,(\not{\!}\!p_{D}+m_{D})/2m_{D}^{1/2},\\
{\P}_{\psi} &=& \not{\!}\!\a_{\psi}\,(\not{\!}\!p_{\psi}+m_{\psi})/2m_{\psi}^{1/2}
\end{eqnarray}

In a more complicated case when we consider a $P$-wave meson like $\chi_c$,
we have to introduce a projector
\begin{equation}
\P_\chi = (\not{\!}\!p_{\bar{c}}-m_c)\not{\!}{\a}_S\,(\not{\!}\!p_c+m_c)/m_\chi^{3/2}
\end{equation}
where the 4-vector $\a_S$ represents the orientation of the quark pair spin momentum
$S$, while the orbital momentum $L$ is related to the quark relative momentum $q$
\begin{equation}
p_{\bar{c}}= p_{\chi}/2 +q, \qquad p_c= p_{\chi}/2 -q,
\end{equation}
as is explained in Refs.~\cite{Krase,Guber}.
The states with definite projections of the spin and orbital momenta $S_z$ and $L_z$
can be translated into states with definite total angular momentum $J_z$ (that is,
the real mesonic states $\chi_{c0}$, $\chi_{c1}$, $\chi_{c2}$) through Clebsch~-~Gordan
coefficients.

The amplitudes for other two-body $W$ decays considered here can be constructed in
a similar manner. The calculation of Feynman diagrams is straightforward and is
performed using the algebraic manipulation system FORM \cite{FORM}.

The formation of the final state mesons can be described in either of the two ways.
In the NRQCD approach, the momenta of the quarks
forming a meson are strictly connected with the meson momenum as
\begin{equation}
p_c= (m_c/m_D)\,p_D;\quad p_s= (m_s/m_D)\,p_D,
\end{equation}
and the identity $m_D=m_c+m_s$ is strictly observed.
The overall probability for forming a bound state is determined by the only parameter,
the radial wave function of a meson at the origin of the coordinate space $\R_{D}(0)$.
\footnote{In view of
the small mass of strange quark, we, strictly speaking, go beyond the range of validity.
Probably, this approach had better be called the NRQCD-ispired or NRQCD-motivated approach. The approach had
been nevertheless used in several researches; such as, for example, in calculating the
fragmentation functions $c\to D_s$ and $c\to D_s^*$ (see Ref.\cite{ref4}). Having this remark
done, we will hereafter refer to this approach as to NRQCD, for the sake of brevity.}

Then, the partial decay widths read
\begin{eqnarray}
\Gamma_{W\to D\,\g} & = & \frac{1}{3}\,
\frac{m_W^2{-}m_D^2}{64\,\pi^2\,m_W^3}\;
\Bigl| \sum_{i=1}^3 \A_i \Bigr|^2 \; \bigl|\R_{D}(0)\bigr|^2,
\end{eqnarray}

\begin{eqnarray}
 \Gamma_{W\to\psi D} & = & \frac{1}{3}\,
 \frac{\lambda^{1/2}(m_W^2,m_\psi^2,m_D^2)}{256\,\pi^3\,m_W^3}
  \Bigl| \sum_{i=1}^5 {\A'}_i \Bigr|^2
  \bigl|\R_{\psi}(0)\bigr|^2\;\bigl|\R_{D}(0)\bigr|^2
\end{eqnarray}

In the Light Cone (LC) approach, the quark momenta can vary,
so that their positive light-cone components $p_i^+ = E_i+ {p_{||}}_i$
are given by
\begin{equation}
p^+_c= z\,p^+_D;\quad p^+_s= (1-z)\,p^+_D
\end{equation}
with $0<z<1$, and the distribution in $z$ is determined by the meson wave function
$\Phi(z)$ (Ref.~\cite{Chernyak}).
The normalization condition is~$\int_0^1 \Phi(z)\,dz~=~1$.
The overall probability for forming a meson is determined by the constant $f_D$
which is related to the NRQCD wave function as
\begin{equation}
|{\R_{D}(0)}|^2/{4\pi}= (m_D/12)\,f_D^2.
\end{equation}

In this approach, the partial decay widths read
\begin{eqnarray}
\Gamma_{W\to D\,\g}  =  \frac{1}{3}\,
\frac{m_W^2{-}m_D^2}{64\,\pi^2\,m_W^3}\;
\Bigl| \sum_{i=1}^3 \I_i \Bigr|^2 \;\frac{m_D\,f_D^{2}}{12},
\end{eqnarray}

\begin{eqnarray}
\Gamma_{W\to\psi D} =  \frac{1}{3}\,
\frac{\lambda^{1/2}(m_W^2,m_\psi^2,m_D^2)}{256\,\pi^3\,m_W^3}
\Bigl| \sum_{i=1}^5 \II_i \Bigr|^2 \frac{m_D\,f_D^{2}\;m_\psi\,f_\psi^{2}}{144},
\end{eqnarray}

where
\begin{eqnarray}
\I_i & = & \int_0^1 \A_i(z)\,\Phi_D(z)\,dz, \label{int1} \\
\II_i & = & \int {\A'}_i(z_1,z_2)\,\Phi_D(z_1)\,\Phi_{\psi}(z_2)\;dz_1\,dz_2
\label{int2}
\end{eqnarray}

The radial wave functions of $J/\psi$, $\psi(2s)$, $\chi_c$, and $B_c$ mesons were
taken from potential models \cite{EiQu2,Quigg}.
Whenever possible, the values of the wave functions were checked for consistency
with the measured decay widths \cite{PDG}.
The radial wave functions of $D_s^{(*)}$ mesons were extracted from the constant $f_D$
shown in Ref.~\cite{PDG}; the latter is close to a theoretical result of Ref.~\cite{RWF1}.

For $B_s^{(*)}$ mesons, we use the value obtained in lattice QCD calculation~\cite{RWF2}.
For illustrative purposes, we take the pseudoscalar and vector wave functions equal,
though theoretically it is not excluded that they may be slightly different~\cite{RWF1,RWF2}.
We have eventually
\begin{eqnarray}
&|\R_{J/\psi}(0)|^2 = 0.80~GeV^3,\nonumber \\
&|\R_{\psi(2s)}(0)|^2 = 0.40~GeV^3,\nonumber \\
&|\R_{\chi_1}'(0)|^2= |\R_{\chi_2}'(0)|^2= 0.075~GeV^5,\label{wave_fun}\\
&|\R_{D_s}(0)|^2=|\R_{D_s^*}(0)|^2= 0.137~GeV^3, \nonumber \\
&|\R_{B_c}(0)|^2=|\R_{B_c^*}(0)|^2= 1.993~GeV^3, \nonumber \\
&|\R_{B_s}(0)|^2=|\R_{B_s^*}(0)|^2= 0.314~GeV^3. \nonumber
\end{eqnarray}

The values of $\R(0)$ and $f$ are not the major source of theoretical uncertainties,
whereas the quark masses and the shapes of $\Phi(z)$ are. We will postpone the discussion
of this issue to the next section.

\section{Results and discussion}
\subsection{Radiative decays $W\to D_s^{(*)}\g$}

\begin{figure}[!ht]
\centerline{\includegraphics[width=4.0in]{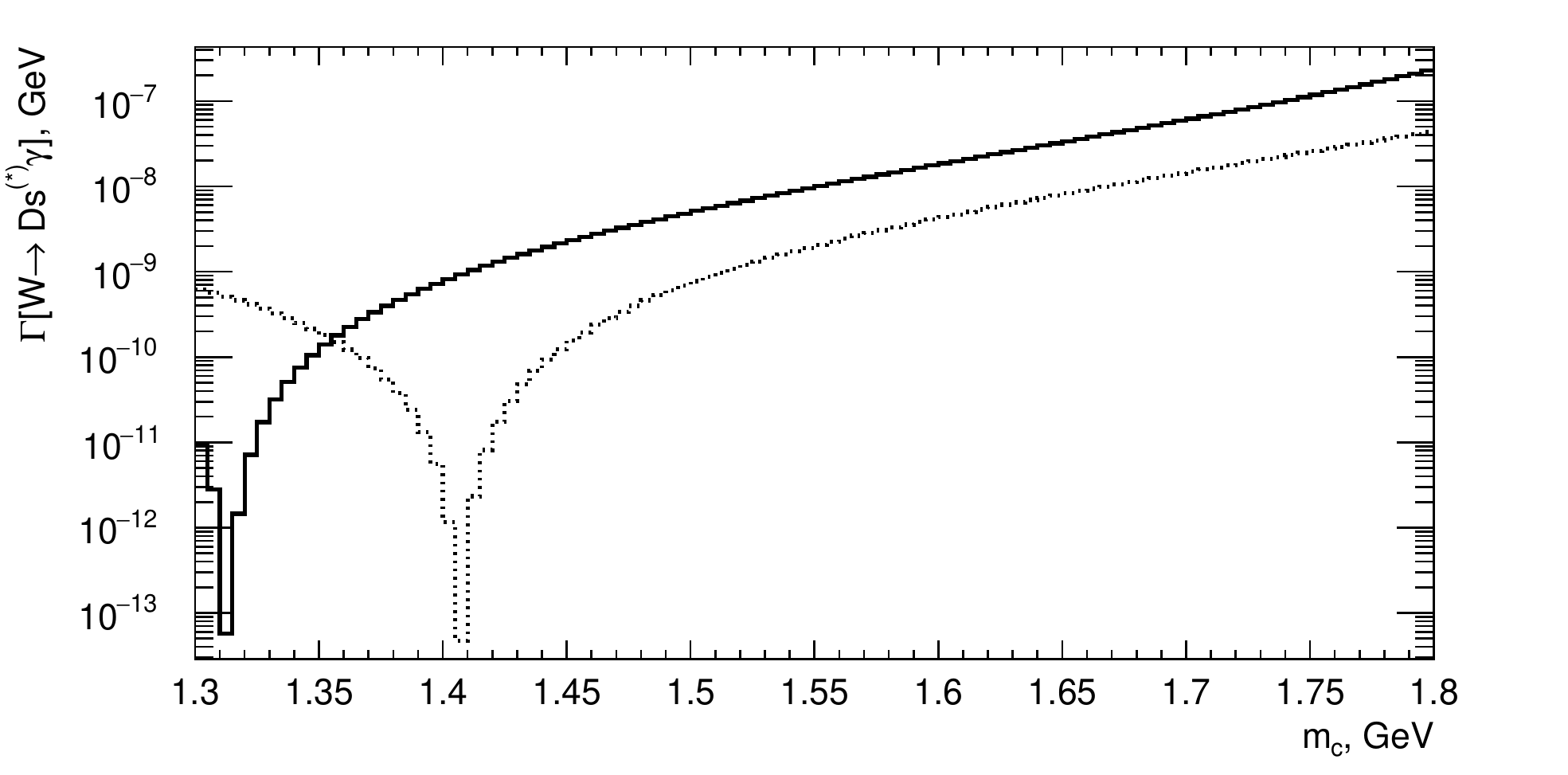}}
\caption{Dependence of the $W\to D_s\g$ decay widths on the quark masses in the
NRQCD scheme. The charmed quark mass $m_c$ is plotted along the x-axis, the strange
quark mass $m_s$ is calculated as $m_s=m_D-m_c$ or $m_s=m_{D^*}-m_c$.
Solid curve, $W\to \g + D_s$;
dotted curve, $W\to \g + {D_s}^*$.
\protect\label{fig:NRQCD_masses}}
\end{figure}

We start the discussion with showing our results for the radiative decays
(\ref{c1}), (\ref{c2}) in the NRQCD scheme. Fig.~\ref{fig:NRQCD_masses} illustrates
the dependence of the predicted decay widths on the choice of quark masses. Recall
that in the model which we are using, the masses of the quarks composing a meson
must strictly sum up to the meson mass. The values of $m_c$ are plotted in
Fig. \ref{fig:NRQCD_masses}
along the x-axis, and then $m_s$ is calculated as $m_s=m_D-m_c$ or $m_s=m_{D^*}-m_c$.

On the other hand, one can argue that at the $W$ scale one should use the pole
masses rather than constituent masses. Then, with setting $m_c = 1.27$ GeV,
$m_s = 93.4$ MeV~\cite{PDG} (and rather unphysical $m_D=m_c+m_s$), we obtain
\begin{equation}
\Gamma(W\to D_s\g) = \Gamma(W\to D_s^*\g) = 1.79\cdot 10^{-9}.
\end{equation}
The sensitivity of the NRQCD results to the quark masses represents our first finding.

Now let us turn to the LC scheme.
Given the fact that the meson energies are much larger than their
masses, one can apparently use the massless approximation, as it is done in refs. \cite{LuchLik,Parsa,Grossman}. However, taking the limit $m_D\to 0$ needs some care.
The physical $D_s^*$ meson is not massless and may have longitudinal polarization.
On the other hand, if we set the meson massless from the very beginning, we are
unable to define its longitudinal polarization vector, and so, are unable to perform
the relevant calculation. This was probably the reason for not showing the respective
results in \cite{Parsa,Grossman}.

In our real calculations we attribute some small but finite values to the quark masses
(while keeping the relation $m_c+m_s=m_D$). By doing this, we obtain a numerically
stable result which is fairly insensitive to the choice of quark masses and their
ratios. Hereafter we will call this the small-mass limit.
An important advantage of using finite (non-zero) masses is that we obtain a numerically
stable result for the longitudinal polarization as well. The latter neither depends on
the quark masses nor on the ratio $m_s/m_c$ and represents in our opinion a physically
consistent description of real massive $D_s$ mesons.
We emphasise that using the strict identity $m_D=0$ would lead to loosing an
essential contribution.

In the small-mass limit, the decay width for the $D^*_s\g$ channel tends to a constant
value equal to that for the $D_s\g$ channel.
The crucial role of the longitudinal polarization is our second finding.

\begin{figure}[!ht]
\centerline{\includegraphics[width=4.0in]{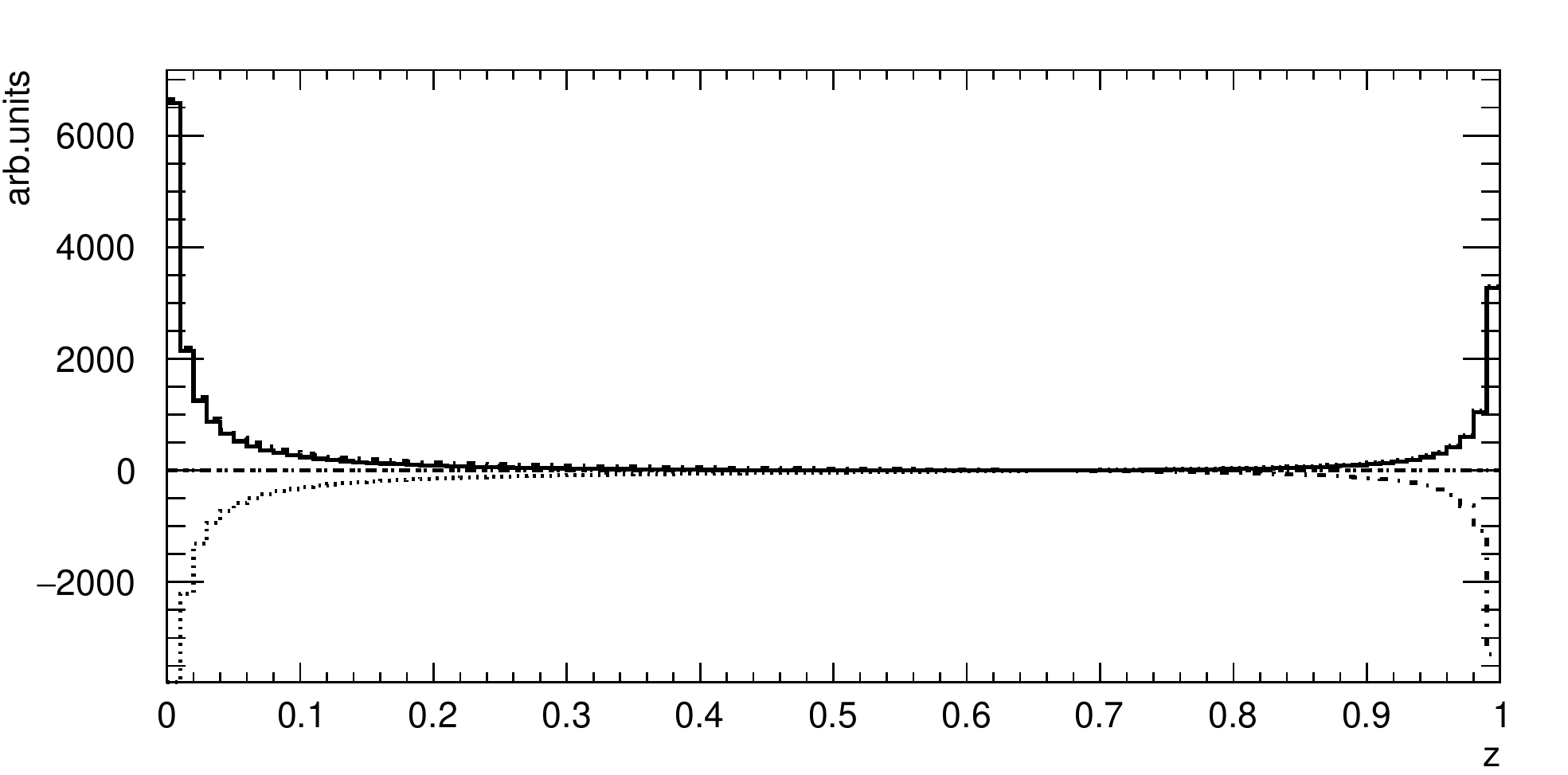}}
\centerline{\includegraphics[width=4.0in]{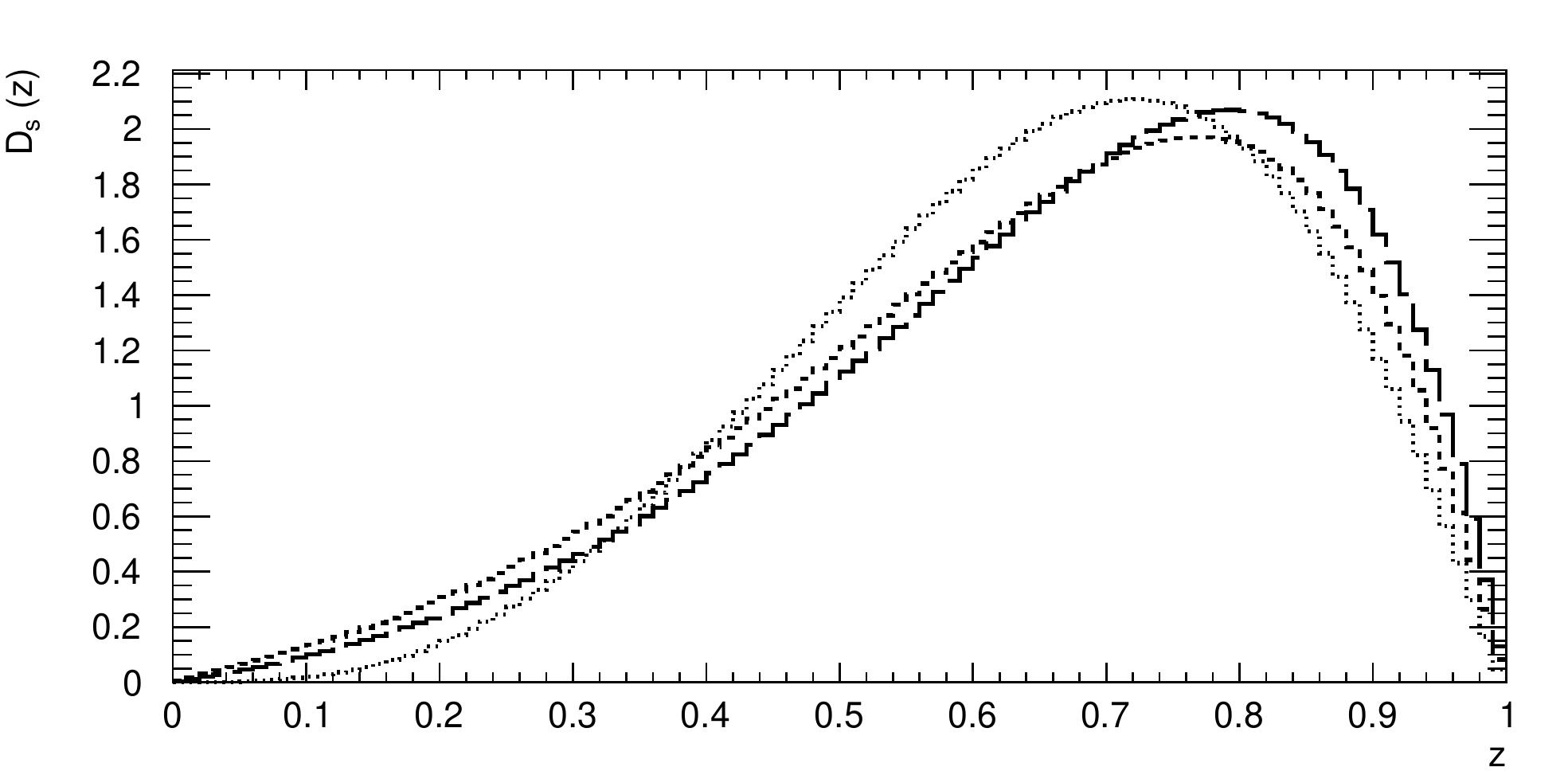}}
\vspace*{8pt}
\caption{%
Upper panel: behavior of the $W\to D_s\g$ decay amplitudes as functions of the
$c$-quark momentum fraction $z$ in the LC scheme. Dashed and dash-dotted curves represent
different polarization states of the $W$ and the photon.\\
Lower panel: examples of different parametrizations of the $D_s$ wave function.
Dotted curve, Equ.~(\ref{Luch}) with $a_c=3.1$, $a_s=1.2$;
dashed curve, Equ.~(\ref{Gros}) with $a_c=0.9$, $a_s=1.1$, $\sigma= 0.279$;
dash-dotted curve, Equ.~(\ref{Gros}) with $a_c=1.1$, $a_s=0.9$, $\sigma= 0.279$.
}
\protect\label{fig:LC_functions}
\end{figure}

An important part of theoretical uncertainties in the LC scheme comes from the shape
of the wave functions. The decay amplitudes (\ref{amp_gamD}) show peaks at $z=0$ and
$z=1$, see Fig. \ref{fig:LC_functions}. That means that the integral (\ref{int1})
is not dominated by the central part of the distribution $\Phi_D(z)$, but rather by
its tails. As a consequence, the functions $\Phi_D(z)$ which look almost
indistinguishable may lead to significantly different predictions for the decay
widths. To illustrate the variability of theoretical predictions we tried the model
parametrizations of the form
\begin{eqnarray}
\Phi_D(z) &\propto& z^{a_c} \bar{z}^{a_s},\label{Gros} \\
\Phi_D(z) &\propto& z^{a_c} \bar{z}^{a_s} \exp(-\bar{z}/\sigma), \label{Luch}
\end{eqnarray}
where $\bar{z}=1-z$, and the respective results are collected in Table~\ref{tab:table0}.
We can conclude that the overall normalization of the wave function is much less
important than its endpoint behavior. This fact constitutes our third finding.

The difference between NRQCD and LC results streams from the fact that NRQCD probes
the central region $z\simeq m_c/m_D$, while LC probes the endpoint regions
$z\simeq 0$ and $z\simeq 1$, which have nothing in common.
Summing up, the overall accuracy of theoretical predictions can hardly be made better
than within one order of magnitude.

\begin{table}[!ht]
\tbl{The $W\to D_s^{(*)}\g$ decay widths calculated for different parametrizations
of the $D_s$ wave functions, eqs. (\ref{Luch}), (\ref{Gros}).}
{\begin{tabularx}{0.95\textwidth}
{c@{\hspace{1.7cm}}c@{\hspace{1.7cm}}c@{\hspace{1.7cm}}c@{\hspace{1.7cm}}c@{\vspace{0.2cm}}}
\toprule
Channel      &$a_c$    & $a_s$  & $\sigma$   &$\Gamma_W$\,Br [GeV] \\[1mm]
\colrule
$W\to D_s^{(*)}\g$     & 3.1     &  1.2   &    --      &    $7.12{\cdot} 10^{-9}$   \\
$W\to D_s^{(*)}\g$     & 1.0     &  1.0   &   0.279    &    $1.96{\cdot} 10^{-8}$   \\
$W\to D_s^{(*)}\g$     & 1.1     &  0.9   &   0.279    &    $1.31{\cdot} 10^{-8}$   \\
$W\to D_s^{(*)}\g$     & 0.9     &  1.1   &   0.279    &    $3.06{\cdot} 10^{-8}$   \\
$W\to D_s^{(*)}\g$     & 1.0     &  1.0   &   0.280    &    $1.94{\cdot} 10^{-8}$   \\
\botrule \\[-3mm]
\end{tabularx}}
\label{tab:table0}
\end{table}

\begin{table}[!ht]
\tbl{Characteristics of the invariant mass distributions for $D_s^{(*)} +\g$
 states produced in different $W$ decay modes.\\[-2mm]}
{\begin{tabularx}{0.95\textwidth}
{l@{\hspace{0.9cm}}c@{\hspace{0.9cm}}c@{\hspace{0.9cm}}c@{\hspace{0.9cm}}c@{\vspace{0.2cm}}}
\toprule
Channel     &$\bar{m}(D_s\g) [GeV]$ & r.m.s. [GeV] &$\Gamma_W$\,Br [GeV] & $\Gamma_W$\,Br [GeV] \\
\!\!\!\!\!  &       &       &NRQCD &LC\\[1mm]
\colrule
$D_s+\g     $ &  80.386  &  1.042  & $9.77{\cdot} 10^{-9}$ & $7.12{\cdot} 10^{-9}$ \\
$D_s^*+\g   $ &  77.749  &  1.837  & $1.86{\cdot} 10^{-9}$ & $7.12{\cdot} 10^{-9}$ \\
\botrule  \\[-3mm]
\end{tabularx}
\label{tab:table3}}
\end{table}

\subsection{Hadronic decays containing $J/\psi$}
Hadronic decays of this kind may proceed both due to strong and electromagnetic
interactions.
The electromagnetic contribution is represented by the amplitudes~(\ref{amp_gamD})
supplemented with a conversion of the photon into $J/\psi$ or $\psi(2S)$ meson
(\ref{ampfactor}). The strong contribution is represented by the amplitudes (\ref{amp_psiD}).
Taken solely, the strong contribution shows almost no dependence on the quark masses.
The behavior of the electromagnetic contribution has been discussed in the previous
subsection.

Our major discovery concerning these decays is that the electromagnetic contribution
is not negligible in comparison with the strong contribution, but even can take over%
\footnote{The electromagnetic contribution was completely ignored in Ref.~\cite{LuchLik}.
At the same time, the authors introduce color octet contribution, which looks misleading.
The color octet production mechanism is incompatible with the definition of exclusive
decay: how would it be possible to change the quantum numbers (the color) of
a $c\bar{c}$ state without passing them to another (new) particle?}.
The relative suppression of the strong contribution comes from the intermediate gluon
propagator.

The case when electromagnetic contributions are comparable with or even larger than
strong contributions is rather rare, though not unique. A similar effect is present
in the decays of Z-boson and H-bosons, see discussion in \cite{ref1,ref2,ref3}.

Our predictions for the NRQCD scheme are shown in Table~\ref{tab:table1}. Shown
there are the mass central value $\bar{m}(\psi\,D_s)$; the dispersion (root of mean
square); and the integral contribution to the $W$ width multiplied by the relevant
branching fractions
($\psi'\to J/\psi\,X,\; \chi_c\to J/\psi\,\gamma,\; D_s^*\to D_s\,X$).
For the processes including photonic contributions (four entries in the bottom part
of the Table~\ref{tab:table1})
the quark mass setting was $m_c=m_\psi/2$=1.55 GeV, $m_s=m_{D^{(*)}} - m_c$.
The exact $\psi(2S)$ mass was, however, taken into account in the amplitude conversion
factor~(\ref{ampfactor}). For all other cases we set $m_c$ to one half of the quarkonium
mass.

\begin{table}[!ht]
\tbl{Characteristics of the invariant mass distributions for $J/\psi+D_s$
 states produced in different $W$ decay modes; NRQCD predictions.}
{\begin{tabularx}{0.95\textwidth}
{l@{\hspace{2cm}}c@{\hspace{1cm}}c@{\hspace{1cm}}c@{\vspace{0.2cm}}}
\toprule
 Channel   &$\bar{m}(\psi{+}D_s)$ [GeV] & r.m.s. [GeV] &$\Gamma_W$\,Br [GeV]\\[1mm]
\colrule
\multicolumn{4}{l}{For strong contributions taken solely} \\
$J/\psi+D_s     $ &      80.35    &   1.04   &   $9.65{\cdot} 10^{-12}$   \\
$J/\psi+D_s^*   $ &      77.73    &   1.84   &   $9.74{\cdot} 10^{-12}$   \\
$\psi(2s)+D_s   $ &      73.75    &   2.63   &   $1.68{\cdot} 10^{-12}$   \\
$\psi(2s)+D_s^* $ &      71.40    &   2.87   &   $1.67{\cdot} 10^{-12}$   \\
$\chi_{c0}+D_s  $ &      75.85    &   2.82   &   $1.96{\cdot} 10^{-13}$   \\
$\chi_{c1}+D_s  $ &      75.85    &   2.82   &   $4.27{\cdot} 10^{-13}$   \\
$\chi_{c2}+D_s  $ &      75.85    &   2.82   &   $8.08{\cdot} 10^{-14}$   \\
$\chi_{c0}+D_s^*$ &      73.38    &   3.09   &   $2.43{\cdot} 10^{-13}$   \\
$\chi_{c1}+D_s^*$ &      73.38    &   3.09   &   $4.25{\cdot} 10^{-13}$   \\
$\chi_{c2}+D_s^*$ &      73.38    &   3.09   &   $8.09{\cdot} 10^{-14}$   \\
\hline
\multicolumn{4}{l}{Strong and photonic contributions taken together including the interference}\\
$J/\psi+D_s     $ &               &          &   $2.09{\cdot} 10^{-11}$     \\
$J/\psi+D_s^*   $ &               &          &   $1.24{\cdot} 10^{-11}$     \\
$\psi(2s)+D_s   $ &               &          &   $5.38{\cdot} 10^{-12}$     \\
$\psi(2s)+D_s^* $ &               &          &   $3.53{\cdot} 10^{-12}$     \\
\botrule \\[-3mm]
\end{tabularx}
\label{tab:table1}}
\end{table}

\begin{figure}[!ht]
\centerline{\includegraphics[width=4.0in]{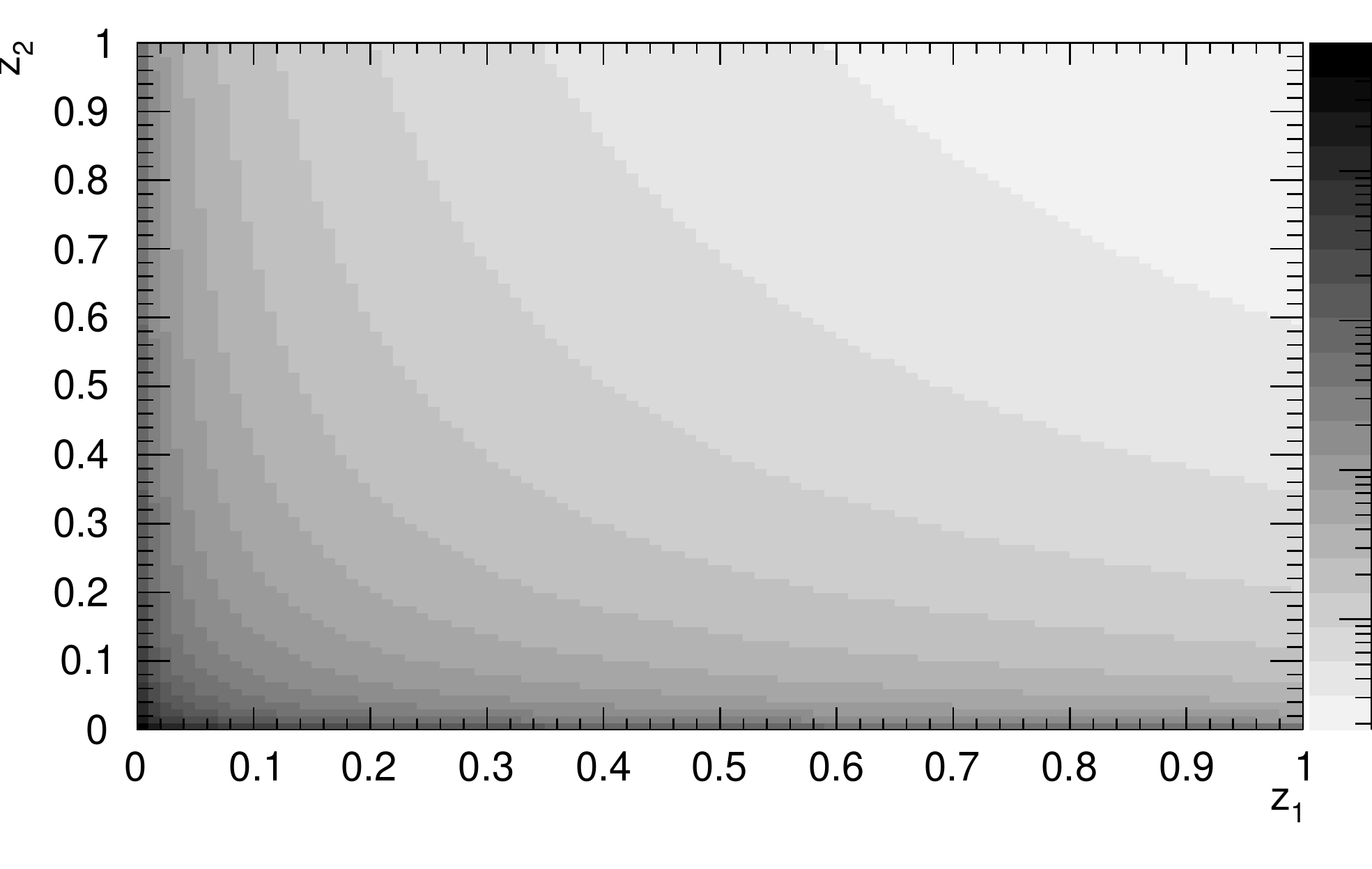}}
\centerline{\includegraphics[width=4.0in]{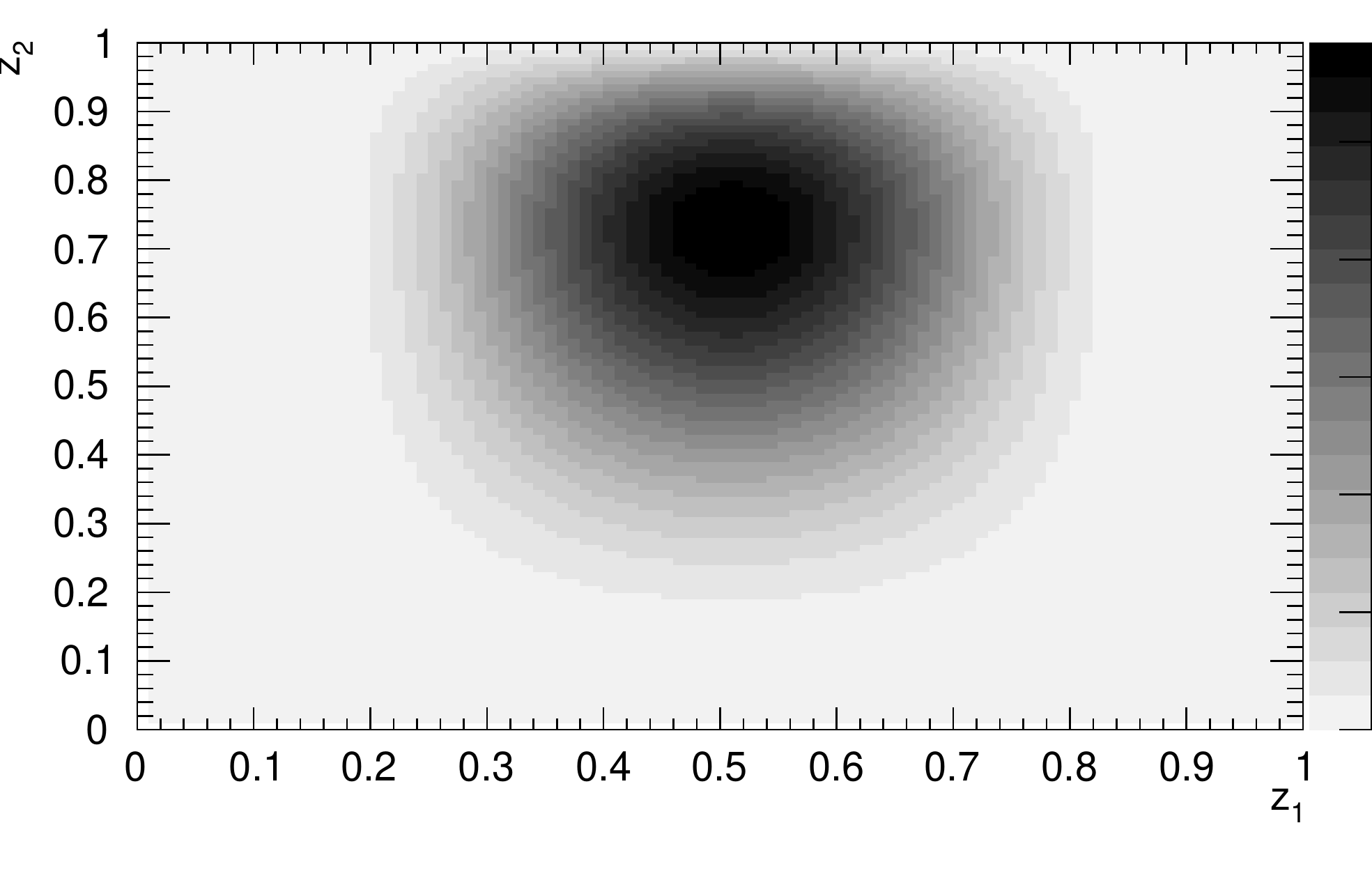}}
\vspace*{8pt}
\caption{%
Upper panel: behavior of the gluon propagator as a function of the
$\bar{c}$-quarks' momentum fractions $z_1$ and $z_2$ in the LC scheme.
The polarization states of the $W$ and the $J/\psi$ are longitudinal.
Note the logarithm scale.\\
Lower panel: behavior of the convolution of $J/\psi$ and $D_s$ wave functions.\\
}
\protect\label{fig:WJD_plots}
\end{figure}

The estimations based on the LC scheme are typically much higher than those based on
NRQCD. This is a consequence of the pole in the gluon propagator in (\ref{amp_psiD}).
This divergence strongly emphasises the region of small quark momentum fractions
(see Fig.~\ref{fig:WJD_plots}) and makes the predictions very sensitive to the endpoint
behavior of the mesons' wave functions. Using the parametrizations proposed in
Ref.~\cite{LuchLik}
\begin{eqnarray}
\Phi_D(z_c) &\propto& (z_c)^{3.1} (z_{\bar{s}})^{1.2}, \n
\Phi_\psi(z_c)&\propto& z_c z_{\bar{c}}\,\exp\{-0.95/(z_c z_{\bar{c}})\}
\end{eqnarray}
with  $z_{\bar{s}}=1{-}z_c$ and $z_{\bar{c}}=1{-}z_c$  we obtain
\begin{eqnarray}
 \Gamma(W\to J\psi D_s) = 3.28{\cdot}10^{-11}\\
\Gamma(W\to J\psi D_s^{*}) = 3.85{\cdot}10^{-11}
\end{eqnarray}

In Table~\ref{tab:tab} we show that the parametrizations of $\Phi_{D}(z_1)$ and
$\Phi_{J/\psi}(z_2)$ may lead to noticeably different predictions for the decay widths.
The distribution amplitudes of $D_s^{(*)}$ mesons are taken according to eqs. (\ref{Gros})
and (\ref{Luch}), and those of the $J/\psi$ meson are taken in the form

\begin{eqnarray}
\Phi_\psi(z)&\propto& z^{b_c}\bar{z}^{b_s}\exp\{-(\bar{z}-0.5)^{2}/\sigma_2\},\label{Gros2} \\
\Phi_\psi(z)&\propto& z^{b_c}\bar{z}^{b_s}\exp\{-0.95/(z\bar{z})\}, \label{Luch2}
\end{eqnarray}

These expressions represent simple low-scale ($\mu^2\simeq m_\psi^2$)
parametrizations of $\Phi(z)$, and our calculations reveal large numerical
uncertainties already at that scale. We use these toy parametrizations for
illustrative purposes, without evolving them to a higher scale $\mu^2\simeq m_W^2$,
because the goal of our study is not in producing 'exact' predictions
(a task not looking feasible in full sense), but rather in giving
an estimate of the overall uncertainty.

We do expect large corrections from Efremov-Radyushkin-Brodsky-Lepage (ERBL)
evolution which can significantly change the behavior of $\Phi(z)$ at $z$
close to 0. At the same time, the evolution in its turn brings additional
uncertainties connected to the accuracy of the ERBL equation at $z\simeq 0$.

We also have to note that the importance of small-$z$ region (which is the region
of zero quark momentum) opens a room for final state interactions. The latter can
hardly be described in a reliable way and make all theoretical predictions even
more uncertain.

\begin{table}[!ht]
\tbl{The resulting decay widths for $W\to J/\psi {D_s}^{(*)}$ decays calculated using different parametrizations of mesons' distribution amplitudes, eqs. (\ref{Luch}), (\ref{Gros}) or (\ref{Luch2}), (\ref{Gros2}).\\}
{\begin{tabularx}
{0.95\textwidth}
{l@{\hspace{0.8cm}}c@{\hspace{0.8cm}}c@{\hspace{0.8cm}}c@{\hspace{0.8cm}}c@{\hspace{0.8cm}}c@{\hspace{0.8cm}}c@{\hspace{0.8cm}}c@{\vspace{0.2cm}}}
\toprule
Channel     &$a_c$ & $a_s$   &$\sigma_1$ &$b_c$  &$b_s$  &$\sigma_2$ &$\Gamma_W$\,Br [GeV] \\[1mm]
\colrule
$J/\psi+D_s$     & 3.1   & 1.2  & --      & 1.0   & 1.0   & --      & $3.28{\cdot} 10^{-11}$ \\
$J/\psi+D_s$     & 1.0   & 1.0  & 0.279   & 1.1   & 0.9   & 0.234   & $6.31{\cdot} 10^{-11}$ \\
$J/\psi+D_s$     & 1.1   & 0.9  & 0.279   & 1.0   & 1.0   & 0.234   & $5.53{\cdot} 10^{-11}$ \\
$J/\psi+D_s^*$   & 3.1   & 1.2  & --      & 1.0   & 1.0   & --      & $3.85{\cdot} 10^{-11}$ \\
$J/\psi+D_s^*$   & 1.0   & 1.0  & 0.279   & 1.1   & 0.9   & 0.234   & $8.44{\cdot} 10^{-11}$ \\
$J/\psi+D_s^*$   & 1.1   & 0.9  & 0.279   & 1.0   & 1.0   & 0.234   & $7.33{\cdot} 10^{-11}$ \\

\botrule \\[-3mm]
\end{tabularx}}
\label{tab:tab}
\end{table}

\subsection{Hadronic decays not containing $J/\psi$}
The channels which do not contain $J/\psi$ meson as an immediate product may still
have it in the final state due to secondary decays:
~$\psi'\to J/\psi\,\pi\,\pi$, ~$\chi_c\to J/\psi\,\gamma$
~$D_s^*\to D_s\,\pi^0$, ~$D_s^*\to D_s\,\gamma$.
The whole group of the decay modes (\ref{d1})-(\ref{d6}) ends up with the
$J/\psi+D_s$ combination, but with distorted invariant mass distribution.

\begin{figure}[!ht]
\centerline{\includegraphics[width=4.0in]{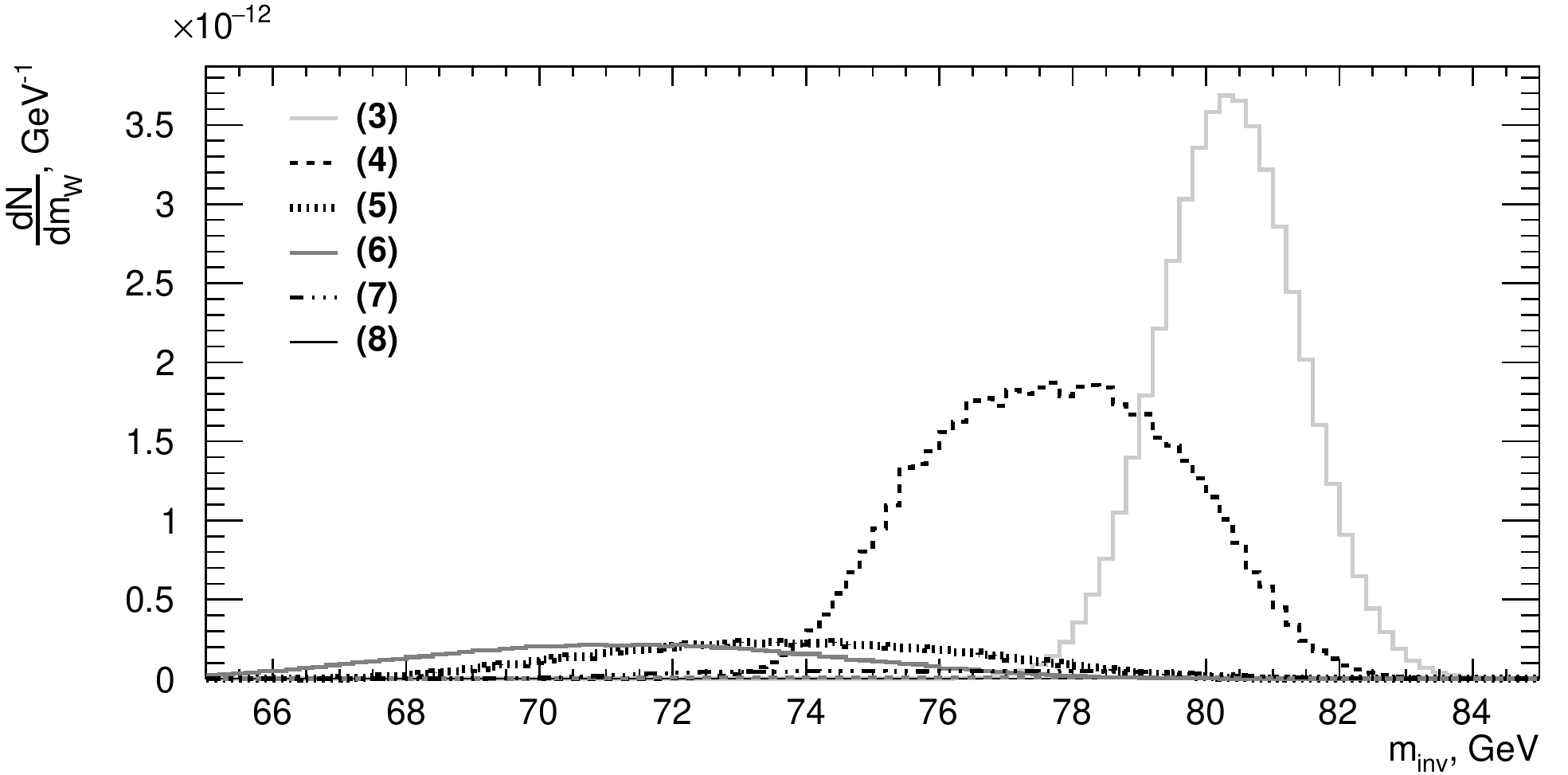}}
\centerline{\includegraphics[width=4.0in]{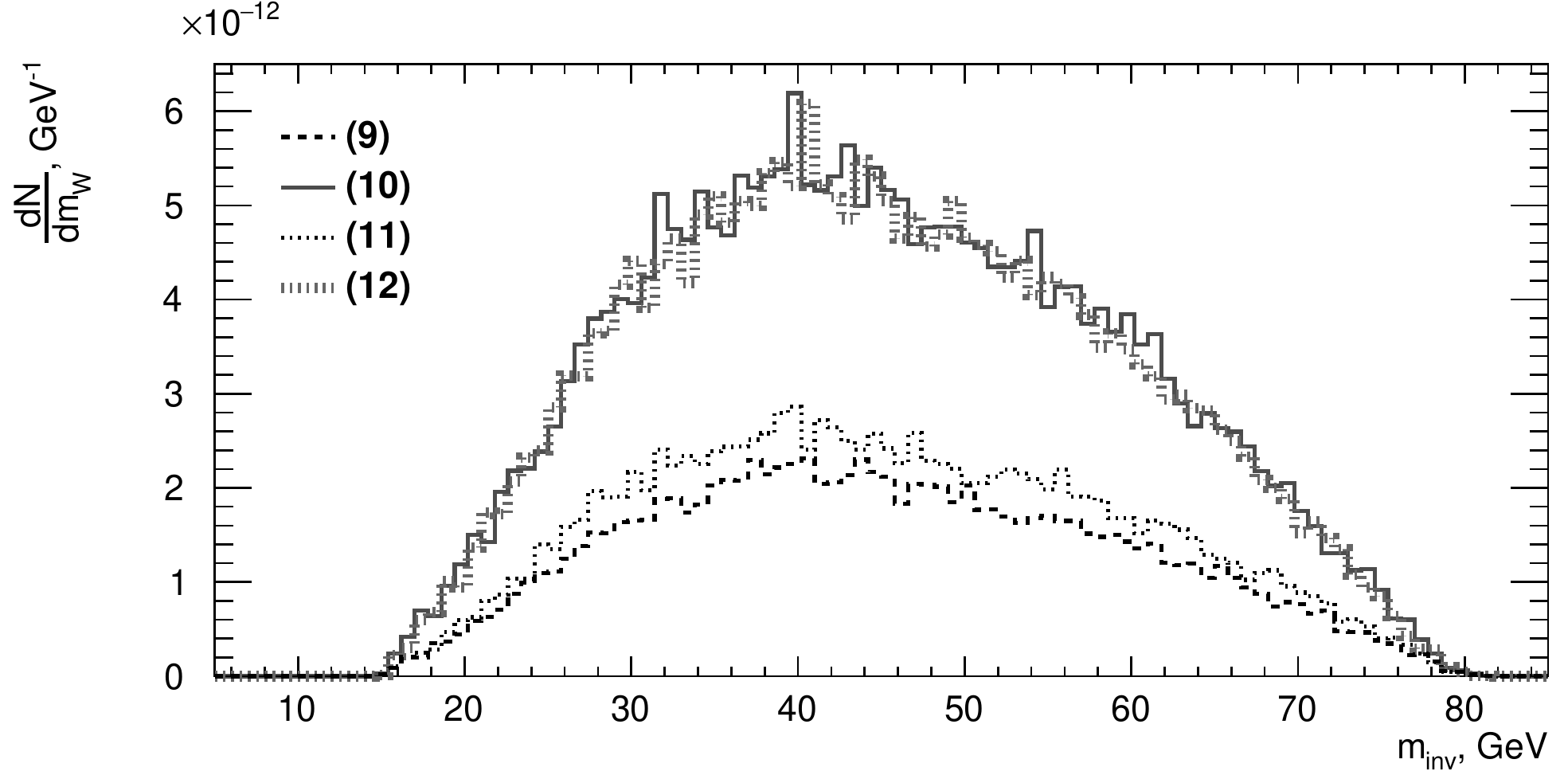}}
\vspace*{8pt}
\caption{The invariant mass distributions for \J +\ds states for different
$W$ decay modes.\\
Upper panel: processes (3) - (8)\\
Lower panel: processes (9) - (12). }
\protect\label{fig:inv_mass}
\end{figure}

Fig.~\ref{fig:inv_mass} displays the invariant mass distributions for $J/\psi\,D_s$
states produced through different $W$ decay modes. The contributions (\ref{d1}) and
(\ref{d2}) overlap and cannot be resolved into individual peaks. This makes the
experimental analysis more complicated and would require a multiparametric (probably,
double-gaussian) fit to describe the signal.
The other contributions (\ref{d3})-(\ref{d6}) are separable from the 'main' peak and
can be rejected by simple kinematic constraints.

In Table~\ref{tab:table2}, we show our results for $b$-flavored modes (\ref{b1})-(\ref{b4}).
The formation of $b$-flavored mesons shows somewhat larger probability
(because of larger values of the wave functions), but these modes may be
less convenient from the point of view of meson identification.

\begin{table}[!ht]
\tbl{Characteristics of the invariant mass distributions for $B_c+B_s$
 states produced in different $W$ decay modes.\\ NRQCD predictions. \\[-2mm]}
{\begin{tabularx}{0.95\textwidth}
{l@{\hspace{1.7cm}}c@{\hspace{1.7cm}}c@{\hspace{1.7cm}}c@{\vspace{0.2cm}}}
\toprule
         &$\bar{m}(B_c{+}B_s)$ [GeV] & r.m.s. [GeV]  &$\Gamma_W$\,Br [GeV]\\
Channel        &            &            &  \\
\colrule
$B_c+B_s     $ &   45.49    &   13.89    &    $5.26{\cdot} 10^{-12}$   \\
$B_c+B_s^*   $ &   45.33    &   13.77    &    $5.73{\cdot} 10^{-12}$   \\
$B_c^*+B_s   $ &   45.41    &   13.78    &    $4.11{\cdot} 10^{-12}$   \\
$B_c^*+B_s^* $ &   45.27    &   13.77    &    $5.55{\cdot} 10^{-12}$   \\
\botrule \\[-3mm]
\end{tabularx}
\label{tab:table2}}
\end{table}

\section{Conclusions}

We have analyzed several exclusive two-body decays of $W$ and reached the following conclusions:

    -- In the NRQCD scheme, the predictions are highly sensitive to the quark mass definition.

    -- In the LC scheme, the predictions are primarily impacted by the behavior of the meson wave function at $z=0$ and $z=1$, with little significance given to its central value.
 This makes the theoretical predictions less certain than it was
previously believed, because of both poorly known behavior of meson wave functions
at the endpoints and the potentially non-negligible role of final state interactions.

    -- The longitudinal polarization of vector mesons dominates the respective W decays, resulting in decay widths equal to those found for pseudoscalar mesons.

Our analysis of the hadronic mode $J/\psi+D_s^{(*)}$ shows that the conclusions derived for the meson-photonic decay modes also hold true. A noteworthy finding is that the electromagnetic contributions to these decays are substantial, and may even surpass the strong contributions.

We also revealed that the secondary decays have a notable impact, leading to overlapping invariant mass distributions for $J/\psi,D_s$ and $J/\psi,D_s^*$ modes. To accurately describe the signal, a multiparametric fitting function is required. The mass distributions for other modes, however, are distinct and separable through kinematic constraints.

The $W$ decay modes involving $b$-flavored mesons exhibit larger branching fractions compared to the $c$-flavored modes. However, the challenge lies in accurately identifying the final mesons in these decays.

Our analysis maintains the previous conclusion~\footnote{If we substitute the
values of the wave functions and coupling constants according to Ref.~\cite{LuchLik},
our results for the decay widths differ by a factor of 3 or even larger, despite the
expressions for the amplitudes look the same.} that the W decay branching fractions
into $J/\psi,D_s^{(*)}$ and $D_s^{(*)},\g$ are still too small to warrant experimental
detection with current sensitivity established at the level $6.5\cdot 10^{-4}$
\cite{lhcbWDsGamma}. The inclusion of new important contributions does not change
this conclusion.

\section*{Acknowledgments}
Alexander Sakharov thanks Prof. Robert Harr from Wayne State University for
initiating fruitful discussions on the topic of the study.
The work of Alsu Bagdatova and Sergey Baranov was
supported by the Russian Science Foundation under Grant No. 22-22-00119.


\end{document}